\documentclass[conference,compsoc]{IEEEtran}
\usepackage{booktabs}
\IEEEoverridecommandlockouts
\usepackage{cite}
\usepackage{multirow} 
\usepackage{amsmath,amssymb,amsfonts}
\usepackage{algorithmic}
\usepackage{graphicx}
\usepackage{enumitem}
\usepackage{textcomp}
\usepackage{xcolor}
\usepackage{url}
\usepackage{multirow}
\def\BibTeX{{\rm B\kern-.05em{\sc i\kern-.025em b}\kern-.08em
    T\kern-.1667em\lower.7ex\hbox{E}\kern-.125emX}}
\begin{document}

\title{\mbox{ProxyPrints}: From Database Breach to Spoof, \\A Plug-and-Play Defense for Biometric Systems}

% \author{
% \IEEEauthorblockN{Yaniv Hacmon, Keren Gorelik, Yisroel Mirsky}
% \IEEEauthorblockA{\textit{Department of Information Systems Engineering} \\
% \textit{Ben-Gurion University of the Negev} \\
% Beer-Sheva, Israel \\
% \{yanivhac, gorelikk, yisroel\}@post.bgu.ac.il}
% \and
% \IEEEauthorblockN{Gilad Gressel}
% \IEEEauthorblockA{\textit{Department of Cyber Security???} \\
% \textit{Amrita Vishwa Vidyapeetham} \\
% Coimbatore, India \\
% gilad.gressel@am.amrita.edu }
% }

\author{
    Yaniv Hacmon\IEEEauthorrefmark{1},
    Keren Gorelik\IEEEauthorrefmark{1},
    Gilad Gressel\IEEEauthorrefmark{2},
    Yisroel Mirsky\IEEEauthorrefmark{1}\\[4pt]
    \IEEEauthorrefmark{1}\textit{Faculty of Computer and Information Science, Ben-Gurion University of the Negev, Beer-Sheva, Israel}\\
    \{yanivhac, gorelikk, yisroel\}@post.bgu.ac.il\\[4pt]
    \IEEEauthorrefmark{2}\textit{Center for Cybersecurity Systems \& Networks, Amrita Vishwa Vidyapeetham, Amritapuri, India}\\
    gilad.gressel@am.amrita.edu
}
% \IEEEauthorblockN{1\textsuperscript{nd} Given Name Surname}
% \and
% \and
% \and
% \and
% \and

%%% PAPER LENGTH FOR ACSAC
% Please ensure that your submission consists of a PDF file of no more than 11 double-column pages, excluding well-marked references and appendices limited to a maximum of 5 pages. The full PDF document must not exceed a total of 16 pages.

\maketitle

\begin{abstract}
Fingerprint recognition systems are widely deployed for authentication and forensic applications, but the security of stored fingerprint data remains a critical vulnerability. While many systems avoid storing raw fingerprint images in favor of minutiae-based templates, recent research shows that these templates can be reverse-engineered to reconstruct realistic fingerprint images, enabling physical spoofing attacks that compromise user identities with no means of remediation.

We present \mbox{ProxyPrints}, the first practical defense that brings cancellable biometrics to existing fingerprint recognition systems without requiring modifications to proprietary matching software. \mbox{ProxyPrints} acts as a transparent middleware layer between the fingerprint scanner and the matching algorithm, transforming each scanned fingerprint into a consistent, unlinkable alias. This transformation allows biometric identities to be revoked and replaced in the event of a breach, without affecting authentication accuracy. Additionally, \mbox{ProxyPrints} provides organizations with breach detection capabilities by enabling the identification of out-of-band spoofing attempts involving compromised aliases.

We evaluate \mbox{ProxyPrints} on standard benchmark datasets and commercial fingerprint recognition systems, demonstrating that it preserves matching performance while offering strong security and revocability. Our open-source implementation includes tools for alias generation and deployment in real-world pipelines, making \mbox{ProxyPrints} a drop-in, scalable solution for fingerprint data protection.
\end{abstract}

\begin{IEEEkeywords}
Fingerprint spoofing, Cancellable biometrics
\end{IEEEkeywords}

\section{Introduction}

Fingerprint biometrics have become a cornerstone of modern authentication and forensic systems, underpinning security across a wide range of sectors, including critical infrastructure, consumer electronics, border control, and financial services \cite{das2018biometrics, rui2018survey, mittal2015fingerprint, roy2017masterprint, marasco2014survey}. Their widespread adoption stems from the distinct advantages of fingerprint patterns: they are unique to each individual, remain stable over time, and are relatively easy to capture. In a typical fingerprint-based authentication system, a user enrolls by scanning their fingerprint, which is then stored by the organization as a reference for verifying future access attempts.

\noindent{\textbf{The Threat.}} However, fingerprints function like lifelong passwords, once compromised, they cannot be changed or revoked. If an adversary obtains a copy of an individual's fingerprint, there is no practical way for the victim to remediate the breach, as one cannot simply "reset" their fingers. Unfortunately, biometric databases are vulnerable to data breaches, as demonstrated in several high-profile incidents~\cite{bbc2015breach, taylor2019major}. In some cases, organizations store fingerprint data as raw images. When such images are stolen, attackers can fabricate physical replicas of the prints . These replicas can then be used to spoof authentication systems by pressing them against fingerprint scanners or even to plant false evidence at crime scenes~\cite{cao2016hacking,matsumoto2002impact,soum2019inkjet,bontrager2018deepmasterprints,kim2019reconstruction}.

\begin{figure}[t]
    \centering
    \includegraphics[width=\columnwidth]{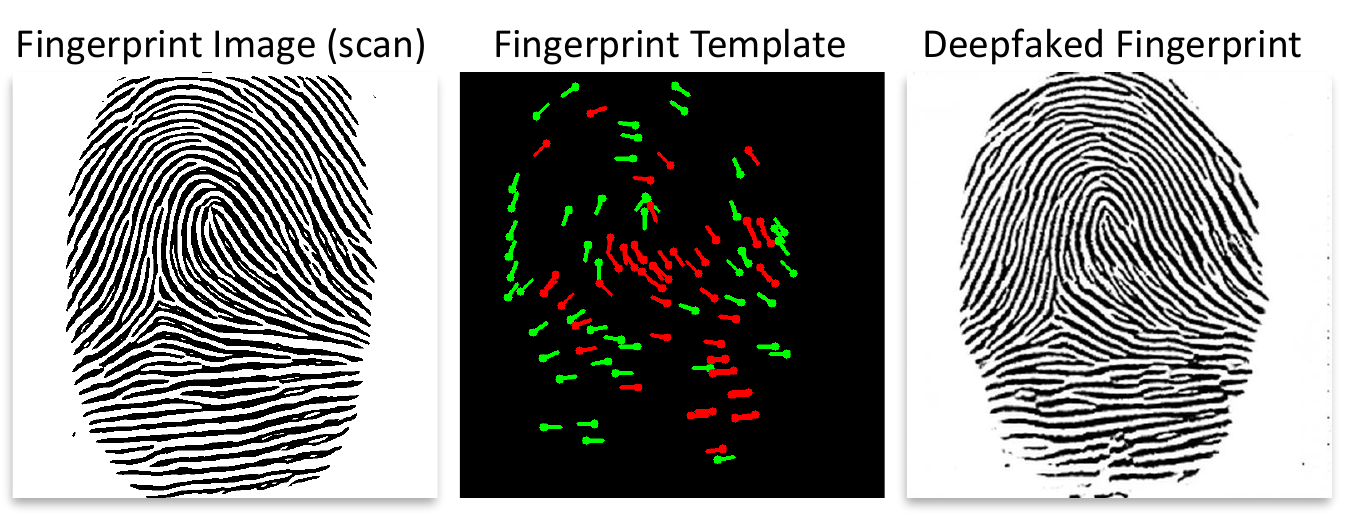}
    \caption{An example demonstrating that templates do not ensure security. Left: The registered fingerprint image. Center: The stored minutiae template. Right: A deepfaked fingerprint image we reconstructed from the template using a custom Pix2Pix model.}
    \label{fig:template}
\end{figure}

Recognizing the risks of storing raw fingerprint images, many organizations instead store fingerprint templates, operating under the assumption that these templates cannot be reverse-engineered to reconstruct the original print. As shown in the center of Fig.~\ref{fig:template}, a fingerprint template encodes the positions and orientations of minutiae distinctive ridge features such as ridge endings and bifurcations. However, this assumption is flawed: templates still retain critical structural information, and recent studies have demonstrated that it is possible to reconstruct high-fidelity fingerprint images from templates, enabling realistic spoofing attacks \cite{cappelli2007fingerprint,kim2019reconstruction,bontrager2018deepmasterprints,moon2021restore}.

\noindent{\textbf{The Research Gap.}} Some efforts have aimed to address this problem through \textit{cancellable biometrics}, where the stored biometric data is a transformed version of the user’s fingerprint. These transformed templates do not directly reveal the original fingerprint but can still be used for matching \cite{marasco2014survey,roy2017masterprint,cappelli2007fingerprint}. However, such approaches typically require overhauling the entire biometric system, as they depend on custom pipelines and proprietary matching algorithms. This is not an option for most organizations since the software is closed source and licensed from a 3rd party vendor. Furthermore, in some cases, the matcher used in some cancellable biometric systems must even be retrained for each individual registration, making these solutions impractical for large-scale deployment \cite{bontrager2018deepmasterprints,kim2019reconstruction,engelsma2022printsgan}.
A further practical limitation of existing cancellable-biometric solutions is their incompatibility with modern cloud-based Automated Biometric Identification Systems (ABIS). Today, many governments and enterprises rely on cloud-hosted fingerprint services such as CloudABIS, Innovatrics, IDEMIA STORM, Biopass ID, and AwareID, where enrollment and matching are performed through remote APIs provided by third-party vendors. Because these systems expose only a standard image or template interface, any defense requiring retraining or modification of the underlying matcher cannot be deployed in such environments. In contrast, a drop-in solution that operates transparently at the image level, without altering the ABIS backend, is essential for practical adoption.

\begin{figure}
    \centering
    \includegraphics[width=\columnwidth]{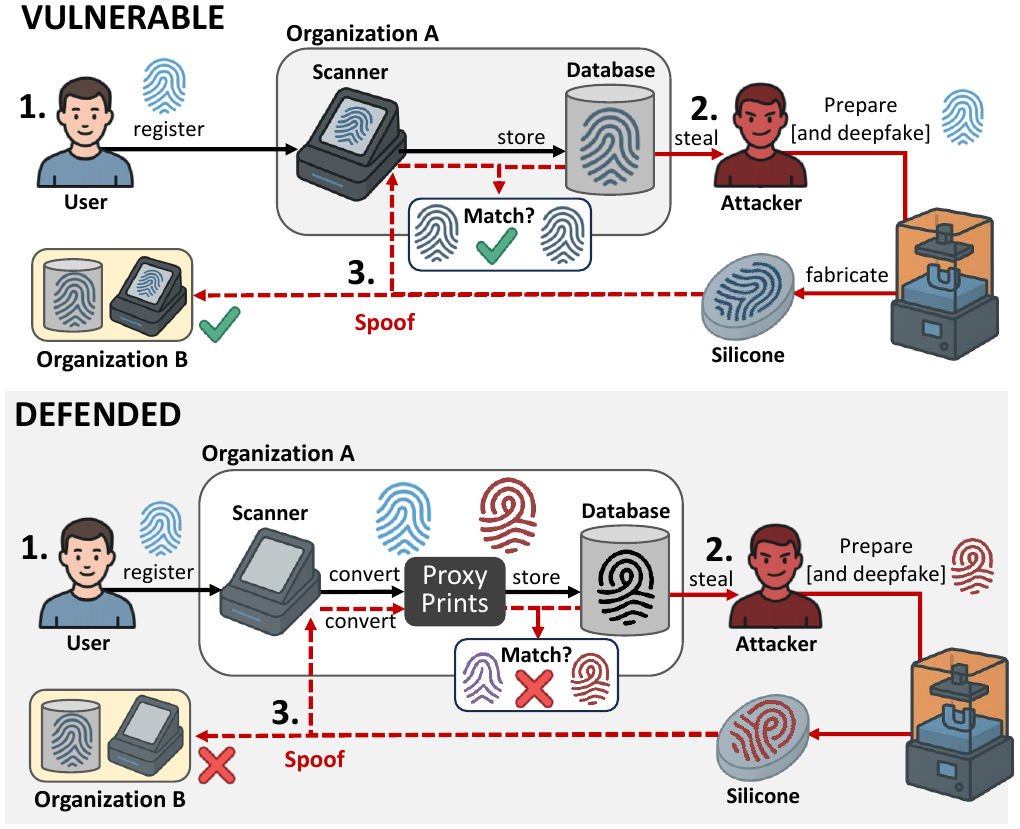}
    \caption{TOP: Illustration of the fingerprint deepfake attack pipeline. An attacker first steals a fingerprint template from a target organization's database. Using a deep generative model, the attacker reconstructs a plausible fingerprint image, which is then 3D printed and cast into a silicone spoof. The spoof is used to unlock fingerprint-based authentication systems by mimicking the legitimate user. BOTTOM: The same attack is thwarted by \mbox{ProxyPrints} as the stolen template will not produce a match.}
    \label{fig:teaser}
\end{figure}

\noindent{\textbf{The Proposed Solution.}} To address these limitations, we propose \mbox{ProxyPrints}, the first plug-and-play module that enables cancellable biometrics in any existing fingerprint recognition system without requiring modifications to the underlying matcher software. As illustrated in Fig.~\ref{fig:teaser}, \mbox{ProxyPrints} operates as an intermediary between the fingerprint scanner and the recognition software, which is often closed-source and provided by third-party vendors. Each scanned fingerprint image is consistently transformed into a new, random identity using a generative AI model. This transformed version is then passed to the recognition system. This transformation effectively assigns each user a persistent biometric alias used for authentication. If the stored alias is compromised, an adversary cannot use it to access this or any other system where the original fingerprint is enrolled because \mbox{ProxyPrints} will produce a fresh alias that differs from the one stored in the database, rendering the attempt invalid. Moreover, \mbox{ProxyPrints} enables organizations to detect potential database breaches, since any use of physical spoofed alias prints outside the system would indicate unauthorized access. Finally, \mbox{ProxyPrints} offers a high level of anonymity since each deployment can be seeded to produce a completely different identity mapping, mitigating potential linkage attacks between separate data breaches. 

\noindent{\textbf{Investigative Study.}} In this work, we also take a critical step beyond prior research. While previous studies have shown that fingerprint templates can be reverse-engineered into fingerprint images, none have conclusively demonstrated whether such reconstructions can succeed in the physical world using real scanners. To validate this threat, we conduct a comprehensive end-to-end evaluation. First, we implemented our own\footnote{We implement our own version because past works have not released functional source code.} template-to-fingerprint deepfake generator using a Pix2Pix-based framework. We then fabricated physical spoofs of the reconstructed fingerprints using a high-resolution resin printer to produce negative molds, followed by silicone casting to replicate the fingerprint surface. Finally, we tested these deepfaked prints across multiple fingerprint scanners and achieved a spoofing success rate of 80\%, at a material cost of just seven cents per spoof. To facilitate reproducibility and further research, we publicly release our deepfake model’s code along with a graphical tool for generating print molds suitable for 3D printing.

% Fingerprint biometrics are a cornerstone of authentication systems and forensics, supporting authentication across sectors including critical infrastructure, consumer electronics, border control, and financial services\cite{das2018biometrics,rui2018survey}. Their widespread adoption is due to the uniqueness, permanence, and ease of acquisition of fingerprint patterns. However, as deployment grows, so do concerns over the security of the stored biometrics. If an  templates, which are increasingly susceptible to misuse and compromise.

\noindent\textbf{Contributions.} In summary, this paper makes the following contributions:
\begin{itemize}
    \item \textbf{\mbox{ProxyPrints}:} We present a practical, drop-in defense that enhances the confidentiality of fingerprint recognition systems without requiring changes to existing (often closed-source) software. The framework not only protect a user's biometrics from data breaches but also enables organizations to detect \textit{when} their database has been breached, enabling immediate remediation.
    
    \item \textbf{Biometric Aliasing Model:} We propose a generative AI model that can be used to create consistent and unlinkable biometric aliases, preserving the accuracy of downstream recognition while protecting the original biometric data. We also show how to seed a model to produce a new mapping \textbf{without any training}.

    \item \textbf{Open-Source Release:} We provide a complete implementation of \mbox{ProxyPrints} online. We also release our template-to-fingerprint deepfake model, along with a graphical tool that automatically generates mesh files for 3D-printable resin molds, enabling reproduction of our spoofing evaluations.
\end{itemize}

% TODO: Add specific numeric results (success rate, quality scores) in abstract and later evaluation section.

% TODO: Replace placeholders with specific citations and references once identified.

\section{Related Work}
\label{sec:related_work}
In this section, we first provide a brief background on fingerprint spoofing. We then review existing biometric protection techniques and compare them to \mbox{ProxyPrints}, highlighting key differences in security, compatibility, and deployability.

\subsection{Fingerprint Synthesis and Spoofing}
The reconstruction and spoofing of fingerprint images from abstract biometric templates have emerged as threats in the field of biometric security. 
Early foundational work by Cappelli et al.~\cite{cappelli2007fingerprint} demonstrated that minutiae templates retain enough information to reconstruct visually plausible fingerprints. This finding highlighted the risk of template inversion even without access to raw biometric data.

The emergence of deep generative models has significantly amplified this threat. Bouzaglo and Keller~\cite{bouzaglo2022synthesis} utilized Generative Adversarial Networks (GANs), specifically a StyleGAN2-based architecture, to synthesize identity-controllable fingerprint images directly from minutiae templates. Kim et al.~\cite{kim2019reconstruction} employed conditional GANs (cGANs) to translate minutiae maps into high-fidelity fingerprint images, achieving match scores sufficient to deceive standard biometric verification systems. Similarly, Moon et al.\cite{moon2021restore} applied a Pix2Pix based image-to-image translation model~\cite{isola2017image}, demonstrating reconstruction from orientation field data and binarized fingerprint skeletons. These approaches collectively show that even lossy or abstract representations of fingerprints can be exploited to regenerate spoof-ready images.

Recent work has also explored multi-identity spoofing using morphing and hybrid generation techniques. For example, Prasad et al.~\cite{bangalore2024gan} proposed a GAN-driven fingerprint morphing strategy that blends features from multiple identities, potentially enabling attackers to create templates that match more than one legitimate user. Such methods highlight the growing sophistication and adaptability of generative models in biometric spoofing scenarios.

% \paragraph{Advanced Defensive Strategies}
% Recent research efforts have expanded beyond traditional cancellable biometrics to explore advanced defensive strategies using deep learning and hybrid systems. Chugh and Jain~\cite{chugh2018spoofbuster} introduced \textit{SpoofBuster}, a convolutional neural network-based system capable of detecting fingerprint spoofs by analyzing local patches around minutiae points, achieving high detection accuracy even against unknown materials. Similarly, Engelsma et al.~\cite{engelsma2022printsgan} proposed a generative adversarial approach that synthesizes fingerprint images while incorporating a discriminator trained to distinguish between real and synthetic patterns, offering insights into spoof detection. Other innovations include dynamic score fusion methods, such as those by Galbally et al.~\cite{galbally2014fingerprint}, which enhance liveness detection by combining multiple sensor signals and image cues. Additionally, Roy et al.~\cite{roy2017masterprint} introduced MasterPrint attacks and concurrently proposed matcher-specific defenses based on analyzing ridge flow inconsistencies. Collectively, these works highlight a growing consensus around combining template transformation, spoof detection, and liveness validation to build more resilient biometric security systems.

\subsection{Cancellable Biometrics and Defensive Methods}

Cancellable biometrics aim to mitigate the irreversibility of biometric leaks by transforming fingerprint templates prior to storage. Early methods, such as block scrambling and geometric warping of ridge maps~\cite{Ratha06,Farooq07}, disrupt minutiae topologies and necessitate either custom matchers or proprietary pre/post-processing, impeding drop-in compatibility. Vector-space techniques like BioHashing and random projections convert minutiae to binary codes matched via Hamming distance~\cite{Teoh04,Nanni06}, but likewise require recompiled recognition software~\cite{Patel15}.

Surveys by Patel et al.\ and Rathgeb and Uhl catalog dozens of such schemes and highlight two recurring limitations~\cite{Patel15,Rathgeb11}. First, most break compatibility with standard Automated Fingerprint Identification System (AFIS) matchers, necessitating retraining or substitution. Second, many are partially invertible or preserve ridge-level cues exploitable for reconstruction~\cite{Xu10,Nasr25}.

Recent deep learning approaches~\cite{Nasr25,Yang21} propose end-to-end cancellable pipelines, but still rely on custom neural matchers and do not integrate with off-the-shelf SDKs. Commercial middleware such as HID BioCore and Secunet Biomiddle focus on device interoperability, not cancelability, and retain untransformed vendor templates~\cite{HIDBioCore24,Secunet23}. MaskPrint~\cite{yan2024maskprint} protects fingerprints via fragmented enrollment across systems, but its partial templates hinder revocation and reuse at scale.

To date, we find no cancellable fingerprint scheme that preserves matcher compatibility, supports transparent middleware deployment, and removes ridge-level invertibility. Existing methods either demand intrusive software changes or leak structural fingerprint information.

\mbox{ProxyPrints} directly addresses these limitations, ensuring compatibility with existing minutiae-based matching infrastructures, introducing high-entropy transformations, and facilitating efficient key rotation. By employing a randomized yet deterministic embedding, \mbox{ProxyPrints} provides the essential properties of security and practical usability that prior cancellable biometric solutions lacked.

\subsection{Fuzzy Extractors and BioHashing}
Fuzzy extractors (FE) and related constructions address the fact that biometric measurements are noisy: two scans of the same finger are close but not identical. The classic FE framework of Dodis et al.\ maps a noisy input $x$ to a uniformly random key $R$ together with public ``helper data'' $P$, such that any $x'$ close to $x$ can reproduce the same $R$ without revealing $x$ \cite{dodis2008fuzzy}. Foundational building blocks include the \emph{fuzzy commitment} \cite{juels1999fuzzy} and \emph{fuzzy vault} \cite{juels2006fuzzy} schemes, which combine error-correcting codes with cryptography to bind secrets to noisy features (e.g., minutiae sets) and later recover them under measurement variation.

BioHashing (BH) takes a different route: a user-specific token (seed) defines a random projection of the biometric feature vector, followed by binarization to yield a fixed-length code matched via Hamming distance \cite{jin2004biohashing}. Because the token is required, BH is often described as ``two-factor'' (biometric + secret) \cite{teoh2006random}.

In practice, both FE and BH produce non-image, code-level representations that are suited for key derivation or code-based matchers. However, they are typically not drop-in compatible with off-the-shelf image/template AFIS pipelines exposed by commercial SDKs or cloud ABIS APIs. Our approach instead keeps the system’s \emph{image} interface intact by performing an image-to-image aliasing transform, while aiming for similar privacy goals (revocability, unlinkability) at the matcher boundary.

% \subsection{Fuzzy Extractors and BioHashing}
% Fuzzy extractors (FE) and BioHashing (BH) address the variability of noisy biometric measurements by mapping an input image $x$ to a stable \emph{code} that can be verified under noise. In FE, helper data and error-correcting mechanisms allow the reconstruction of a cryptographic key from noisy inputs without revealing the biometric itself; in BH, the biometric feature vector is projected (often via a user-specific random matrix) and then binarized to form a code for fast matching. These approaches are effective for key derivation and privacy because the resulting representation is a non-invertible code rather than an image. However, their outputs are not image-compatible and thus typically require code-based matchers or verifier-specific pipelines, reducing drop-in compatibility with off-the-shelf image-based AFIS systems. In contrast, our approach focuses on image-to-image transformations to maintain compatibility with existing matchers while aiming for privacy via aliasing.

\section{Notation \& Threat Model}
\label{sec:thread_model}
In this section we present the notation used across the paper and define our assumed threat model.

\subsection{Notation}
Throughout this paper, we denote a fingerprint template as $t$, which consists of a structured list of minutiae points capturing the fingerprint’s unique features. Its corresponding image-based representation is written as $x_t$, while the original fingerprint image is denoted $x_p$ (as shown in the center and left of Fig.~\ref{fig:template}, respectively).

To refer to specific impressions, we use superscripts: $x_t^i$ and $x_p^i$ represent the template image and genuine fingerprint image for the $i$-th impression. Variations between impressions of the same finger (i.e., different values of $i$) arise naturally from differences in pressure, orientation, or skin deformation during capture.

We define the generative model as a function $G$, where $G(x_t^i) = \hat{x}_p^i$. That is, given a template image $x_t^i$, the model synthesizes a fingerprint image $\hat{x}_p^i$ that closely resembles the genuine impression $x_p^i$.

\subsection{Threat Model}
\label{subsec:threat_model}
We consider a threat model in which the adversary's goal is either: (1) to gain unauthorized access to protected systems by presenting spoofed fingerprint biometrics, or (2) to fabricate synthetic forensic evidence (e.g., latent fingerprints) at crime scenes, thereby falsely implicating or coercing individuals.

\noindent\textbf{Template Acquisition.} 
The adversary’s first step is to obtain a fingerprint template $t$ belonging to the target victim. We assume this can be achieved through two primary vectors:
(1) acquiring leaked biometric templates from illicit online marketplaces , or
(2) compromising biometric databases via targeted intrusions against organizations that store such data\cite{bbc2015breach, taylor2019major}.

\noindent\textbf{Deepfake Fingerprint Synthesis.} 
Given access to $t$, the attacker generates a visual fingerprint image $x_t$ from the template and then applies an image-to-image generative model $f$ to synthesize a realistic fingerprint image $\hat{x}_p$ that resembles the victim’s actual fingerprint $x_p$. To train $f$, the attacker leverages publicly available fingerprint datasets, such as those in \cite{maio2002fvc2002, cappelli2000synthetic}, which contain real or synthetic fingerprint images. Templates can be extracted from these datasets using standard fingerprint SDKs, including NBIS \cite{nist2010nbis}, VeriFinger \cite{verifinger2022}, or SourceAFIS \cite{galuska2015sourceafis}. The resulting training data is denoted as $(x_t^i, x_p^i) \in \mathcal{D}_{\text{train}}$.

\noindent\textbf{Physical Spoof Fabrication.} 
Once the synthetic fingerprint $\hat{x}_p$ is created, the attacker fabricates a physical artifact capable of deceiving commercial biometric readers. This is typically accomplished by producing a silicone replica whose ridges match those in $\hat{x}_p$. When applied to a surface (e.g., a fingerprint scanner), this replica can leave an impression or be directly sensed. While many modern scanners employ liveness detection mechanisms to thwart spoofing attempts, prior work has demonstrated multiple evasion techniques \cite{chugh2018spoofbuster, marasco2014survey, komulainen2015face}.

% In our evaluation, we target commercially available fingerprint readers, including one model equipped with proprietary Fake Finger Rejection Technology[]. To circumvent this defense, we employ thin silicone films that retain the ridge structure of $\hat{x}_p$ while allowing the natural capacitance and optical properties of the attacker’s finger to transmit through the material, thereby bypassing liveness detection.

In our evaluation, we target commercially available fingerprint readers. To simulate realistic spoofing attacks, we fabricate thin silicone films that preserve the ridge detail of $\hat{x}_p$ while remaining permeable to the natural capacitance and optical properties of the underlying fingertip. This design allows the spoof to mimic the physical characteristics of live skin, giving potential to evade standard liveness detection mechanisms.

% TODO: Provide more detailed configuration and training parameters in the Evaluation section.

\section{Threat Validation}

As previously discussed, many organizations opt to store fingerprint templates instead of raw images, assuming this provides protection against spoofing in the event of a data breach. This section demonstrates that this assumption is flawed: a stolen fingerprint template can be reliably converted into a realistic fingerprint image, which, once physically fabricated, can be used to successfully authenticate against commercial fingerprint recognition systems.

\subsection{Deepfake: From Template ($t$) to Fingerprint Image ($x_p$)}\label{sec:method}
While prior works propose various models for generating fingerprint images from templates, we were unable to reproduce their results. Since our goal is not to innovate on synthesis methods but to enable physical realization for security evaluation, we developed a custom approach using a well-established generative AI framework.

We frame fingerprint synthesis as an image-to-image translation task, where a template image $x_t$ is transformed into a realistic fingerprint $\hat{x}_p$. Therefore, we use Pix2Pix~\cite{isola2017image} which is suited for paired image translation tasks.

\subsubsection{Pix2Pix Architecture}\label{subsec:model}
\begin{figure}[t]
    \centering
    \includegraphics[width=\columnwidth]{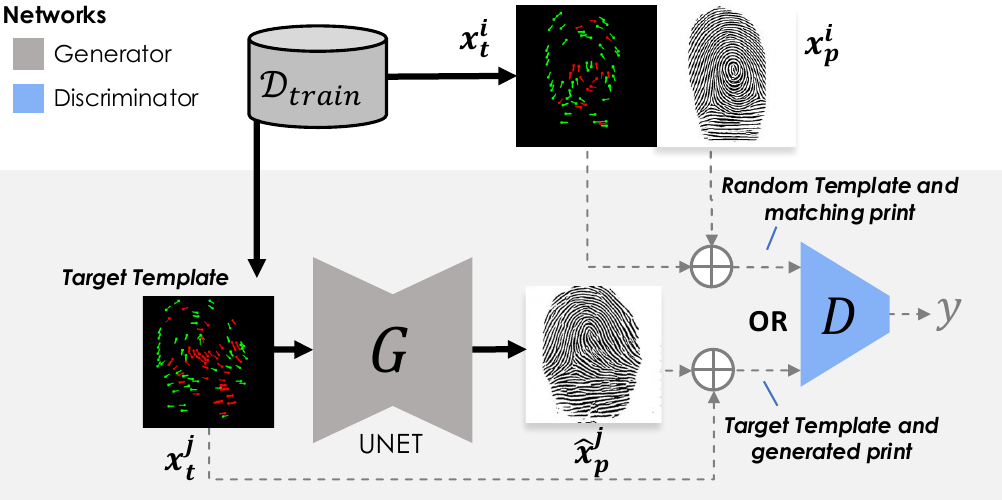}
    \caption{The Pix2Pix framework used to perform the fingerprint template to fingerprint image translation task.}
    \label{fig:arch}
\end{figure}

Pix2Pix consists of a conditional GAN architecture, consisting of a Generator $G$ (U-Net), that maps $x_t \rightarrow \hat{x}_p$, and a Discriminator $D$ which distinguishes real from generated image pairs. A sketch of the architecture can be found in Fig. \ref{fig:arch}.

The adversarial loss is:

$$
\mathcal{L}_{adv} = \mathbb{E}_{x_t, x_p}[\log D(x_t, x_p)] + \mathbb{E}_{x_t}[\log(1 - D(x_t, G(x_t)))]
$$

To enforce structural fidelity, we add an L1 loss:

$$
\mathcal{L}_{L1} = \mathbb{E}_{x_t, x_p}[ \| x_p - G(x_t) \|_1 ]
$$

The final objective:

$$
G^* = \arg \min_G \max_D \mathcal{L}_{adv} + \lambda \mathcal{L}_{L1}
$$

During inference, only $G$ is used to synthesize realistic fingerprints from templates.

\subsubsection{Feature Representation}\label{subsec:feature}
Fingerprint templates are abstract minutiae data, not images. To adapt them for image translation, we encode each minutiae point as a dot with an orientation line. Following \cite{bouzaglo2022synthesis}, we use the red channel to indicate bifurcation points (ridge splits), and the green channel to capture termination points (ridge endings).
This encoding yields a 2D RGB template image $x_t$, which Pix2Pix translates into a synthetic fingerprint $\hat{x}_p$.

\begin{figure}[t]
    \centering
    \includegraphics[width=\columnwidth]{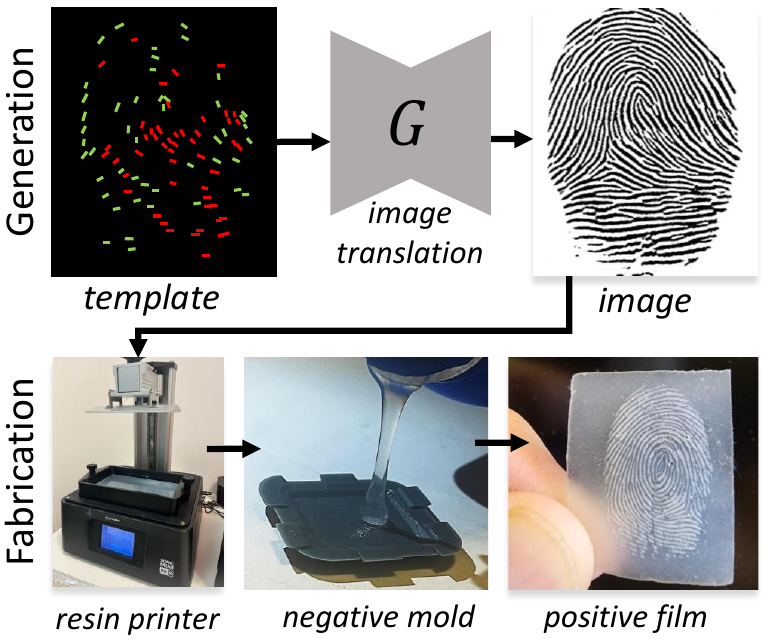}
    \caption{The complete attack process: (1) a stolen fingerprint template is used to generate a matching fingerprint image, (2) a 3D mesh mold is created, (3) the mold is printed using a high-resolution resin printer, (4) silicone is poured into the mold and cures, (5) the resulting spoof is used to deceive biometric scanners.}
    \label{fig:setup}
\end{figure}

\subsection{Fabrication: From Fingerprint Image to Physical Spoof.}
To create a physical spoof of the print, we materialized a fingerprint as a thin silicone film, which acts as a synthetic skin layer that can be covertly worn on a fingertip. We established a reliable three-step process:

\begin{description}
\item[1. Design 3D Mold.]
We converted each 2D fingerprint image into a 3D mesh representing a negative mold, where ridges appear as valleys. This ensures the resulting silicone film has raised ridges matching the original fingerprint. Molds are shallow plates with raised borders to contain resin. We developed an open-source tool with a GUI for generating printable STL files, customizable for ridge depth and mold dimensions.

\item[2. Print Mold with Resin Printer.]  
Molds were printed using a high-resolution Phrozen Sonic Mini 8K S resin printer (22-micron resolution, \$400) and Aqua 8K resin (\$39/kg). Resin printing is essential to capture fine ridge detail (100–300 microns). Each batch of molds took on average thirty minutes to print, five minutes to wash, and an hour to cure.

\item[3. Cast Silicone Film.]  
A two-part silicone (Siraya Tech Defiant 25, \$30/kg) was poured into the mold and cured for three hours. It avoids cure inhibition on resin surfaces, producing fully set, clean films.
\end{description}

Excluding the printer, the cost per fingerprint replica is approximately \$0.07, making this attack method low-cost and highly feasible.

\subsection{End-to-end Attack Evaluation}

\noindent\textbf{Deepfake Model Training.} To train the Pix2Pix model, we utilized the \textbf{LivDet} datasets from 2009, 2011, 2013, and 2015 \cite{galbally2009livdet2009, yambay2012livdet2011, ghiani2013livdet, mura2015livdet2015}, which include live and spoof fingerprint samples collected under varied acquisition conditions.
The dataset was split 80/20 for training and testing, ensuring identity exclusivity across splits to prevent overfitting and identity leakage. Preprocessing involved contour-based cropping, NFIQ2-based quality filtering (retaining only samples with scores $\geq$ 3), standardizing all images to $256 \times 256$ grayscale PNGs, and enhancement. Minutiae points were extracted using \texttt{mindtct} from NBIS and converted to image-based representations $x_t$.
The model was trained with a batch size of 128 over 300 epochs using a learning rate of 0.0002 with the Adam optimizer ($b_1 = 0.5$, $b_2 = 0.999$). We used linear decay on the learning rate after 150 epochs. The complete configuration of our model can be found with our published source code.

\noindent\textbf{Experiment Setup.}
Once the model was trained, we selected 10 fingerprint templates from the test set, and used the model to generate 10 corresponding deepfaked fingerprint images. The templates were selected randomly, with the  criterion that each selected print had a minimum NFIQ-2 score of 50 to ensure a baseline level of quality. Furthermore, we ensured that each fingerprint template originated from a unique identity to avoid identity-based bias in the generated outputs.

We evaluated the 10 films for the 10 fingerprints by authenticating on the ZKTeco Live 20R and SecuGen HU20 (Hamster Pro 20) scanners. For each scanner, we registered the original fingerprint first before testing the film-based replica. Each fingerprint was applied 10 times and the performance of the attack was measured. We used the Bozorth3 algorithm from the NIST Biometric Image Software (NBIS) suite \cite{ko2007user} for matching. Bozorth3 scores typically range from 0 to over 900, with scores above 40 generally indicating a true match\cite{ko2007user}.

\begin{figure}
    \centering
    \includegraphics[width=\columnwidth]{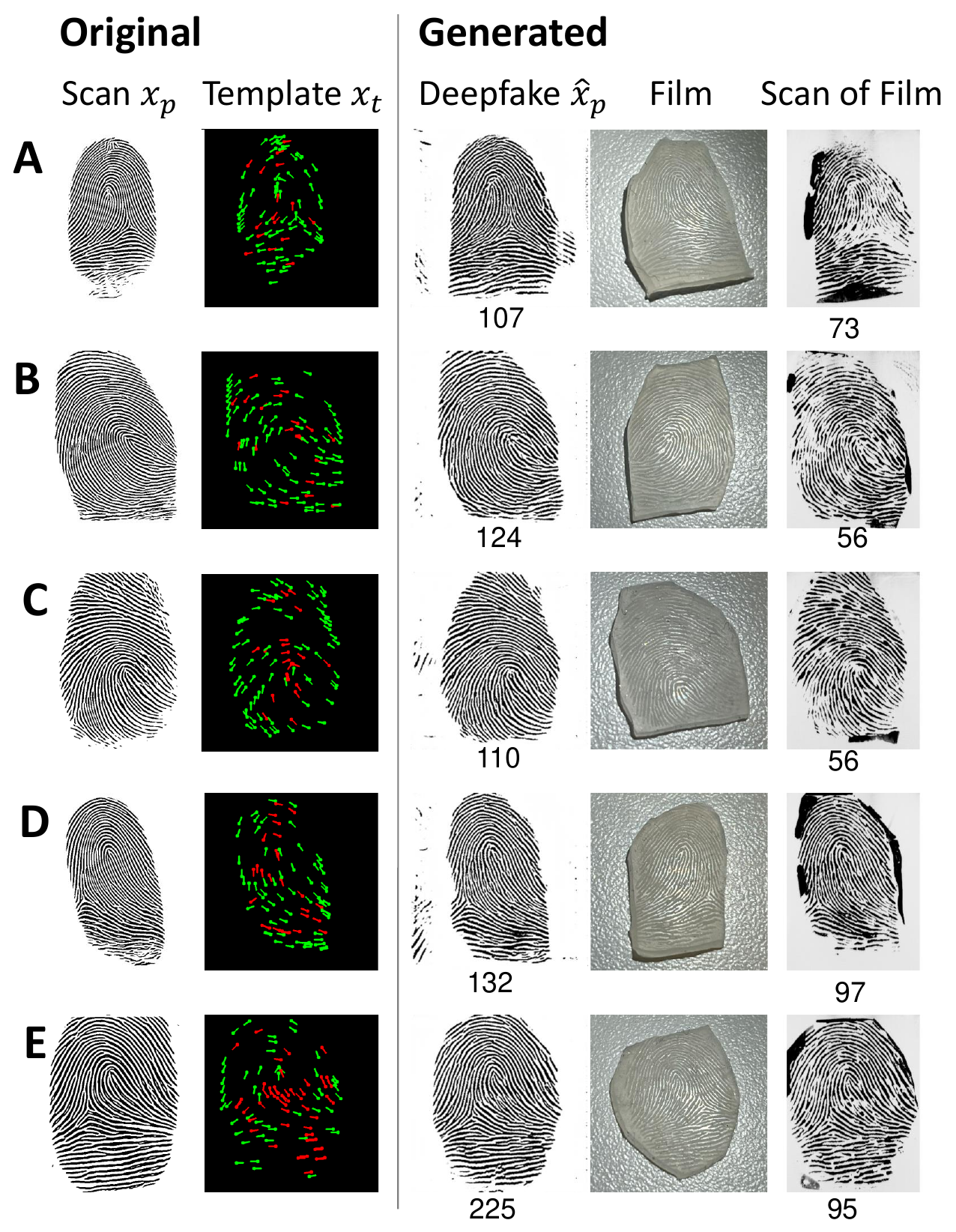}
    \vspace{-1.5em}
    \caption{A sample of the fingerprints used in the evaluation. From left to right: the original fingerprint, its template, the reconstructed fingerprint (deepfake), the silicone film of the fingerprint, the scan of the deepfake fingerprint. The numbers are the Bozorth3 match scores of each fingerprint image to the original $x_p$}
    \label{fig:deepfakes}
        \vspace{-1em}
\end{figure}

\begin{table}[t]
\centering
\caption{Comparison of Fingerprint Scanner Performance Metrics across Samples A–J}
\renewcommand{\arraystretch}{0.8}  % Decrease row height
\begin{tabular}{l|l@{\hskip 2pt}c@{\hskip 2pt}c@{\hskip 2pt}c@{\hskip 2pt}c@{\hskip 2pt}c@{\hskip 2pt}c@{\hskip 2pt}c@{\hskip 2pt}c@{\hskip 2pt}c@{\hskip 2pt}c}
\toprule
\textbf{Scanner} & \textbf{Metric}       & \textbf{A} & \textbf{B} & \textbf{C} & \textbf{D} & \textbf{E} & \textbf{F} & \textbf{G} & \textbf{H} & \textbf{I} & \textbf{J} \\
\midrule
\multirow{4}{*}{SecuGen} 
& Minutiae \#    & 94  & 115 & 91  & 94  & 119 & 115 & 120 & 111 & 141 & 112 \\
& NFIQ2          & 57  & 68  & 63  & 62  & 67  & 65  & 70  & 65  & 21  & 62  \\
& Match Score    & 56  & 50  & 43  & 63  & 64  & 60  & 51  & 61  & 20  & 47  \\
& Success Rate   & \(\frac{10}{10}\) & \(\frac{10}{10}\) & \(\frac{7}{10}\) & \(\frac{10}{10}\) & \(\frac{10}{10}\) & \(\frac{10}{10}\) & \(\frac{8}{10}\) & \(\frac{8}{10}\) & \(\frac{0}{10}\) & \(\frac{7}{10}\) \\
\midrule
\multirow{4}{*}{ZKTeco} 
& Minutiae \#    & 110 & 107 & 103 & 95  & 99  & 130 & 105 & 101 & 116 & 110 \\
& NFIQ2          & 49  & 77  & 72  & 56  & 67  & 48  & 77  & 58  & 22  & 60  \\
& Match Score    & 59  & 59  & 47  & 63  & 52  & 48  & 62  & 42  & 20  & 18  \\
& Success Rate   & \(\frac{10}{10}\) & \(\frac{10}{10}\) & \(\frac{8}{10}\) & \(\frac{10}{10}\) & \(\frac{10}{10}\) & \(\frac{8}{10}\) & \(\frac{9}{10}\) & \(\frac{5}{10}\) & \(\frac{0}{10}\) & \(\frac{0}{10}\) \\
\bottomrule
\end{tabular}
\label{tab:scanner_comparison}
\end{table}

\begin{table}[t]
\centering
\caption{Generated vs Original Fingerprint Metrics (Average)}
\label{tab:generated_vs_original}
\resizebox{\columnwidth}{!}{%
\begin{tabular}{lccc}
\hline
 & \textbf{Match}  & \textbf{Minutiae}  & \textbf{Quality}\\
\textbf{Fingerprint Image} &  \textbf{Score} &  \textbf{Count} & \textbf{(NFIQ2)}\\ 
\hline
Original scan ($x_p$) & 591 & 88.6 & 63.8 \\ 
Deepfake before materialization ($\hat{x}_p$) & 116.6 & 117.8 & 69.2 \\ 
Deepfake after materialization & 51.96 & 111.22 & 60.54 \\ 
\hline
\end{tabular}
}
\end{table}

\noindent\textbf{Results.} Figure~\ref{fig:deepfakes} illustrates the attack pipeline for 5 of the 10 identities (A–E), from the original fingerprint \( x_p \) to the final rescanned silicone spoof. The high visual similarity across each stage confirms the preservation of identity-specific features, including ridge flow and minutiae patterns, despite multiple transformations (template conversion, generation, 3D printing, and silicone casting). These findings support the central claim that abstract templates \( t \) can be inverted into operational spoof artifacts.

Table~\ref{tab:generated_vs_original} summarizes average biometric and image quality metrics across original, generated, and physical spoofed fingerprints. Despite degradation during materialization, spoofed fingerprints maintained a Bozorth3 match score of 51.96, above the common authentication threshold of 40, proving attack feasibility. Interestingly, deepfakes before materialization scored higher (116.6), benefiting from the clean generation process.

While minutiae count increased in synthetic prints (due to generation or casting artifacts), this did not impact successful matching. NFIQ2 quality remained high across all variants, with generated images scoring best, likely due to enhanced fingerprint training data. These results indicate both high visual realism and functional biometric similarity.

While the aggregate statistics provide a strong overall signal, it is also valuable to examine performance at the individual impression level. Table~\ref{tab:scanner_comparison} presents the per-template attack success rate, average match score, and average quality score for each of the ten evaluated impressions. In nearly all cases, the spoofed fingerprints were consistently accepted by the scanners across repeated trials, with success rates approaching or achieving $10/10$. This indicates that the physical replicas are not only theoretically plausible but practically deployable, even under repeated interaction with commercial fingerprint readers.

% However, two cases, impressions $D$ and $E$, notably underperformed. As can be seen in Fig.~\ref{fig:deepfakes}, visual inspection of the final spoof scans for these impressions reveals imperfections in the ridge structure and global alignment. Upon closer examination, these issues appear to stem from defects introduced during mold printing or silicone casting. In particular, uneven resin exposure or warping during curing may have introduced fine-scale distortions that disrupted the fidelity of the ridge topology.

It is important to contextualize the observed failures (of identities I and J) within the broader threat model. An adversary in possession of the stolen template \( t \) can test fabricated spoofs against the synthetic fingerprint \( \hat{x}_p \) using their own scanner and equipment prior to launching the attack. This capability allows the attacker to iterate on mold quality at no operational risk, retrying as many times as needed until a sufficiently high-quality replica is produced. As such, while certain impressions may initially fail due to fabrication noise, the attacker can feasibly ensure high-probability success in practice.

\noindent\textbf{Summary.} Our experiment demonstrates that deepfaked fingerprints can successfully unlock \textbf{real-world} biometric scanners. This highlights that not only stored fingerprint images but also stored fingerprint templates pose a serious risk to user privacy and security if breached.

% \begin{figure}[t]
%     \centering
%     \includegraphics[width=1\linewidth]{Fig/Data_Process.png}
%     \caption{An example of the preprocessing pipeline, showing the transformation from a raw fingerprint scan to its cropped, centered, and enhanced representation.}
%     \label{fig:preprocessing}
% \end{figure}

\begin{figure}
    \centering
    \includegraphics[width=.7\linewidth]{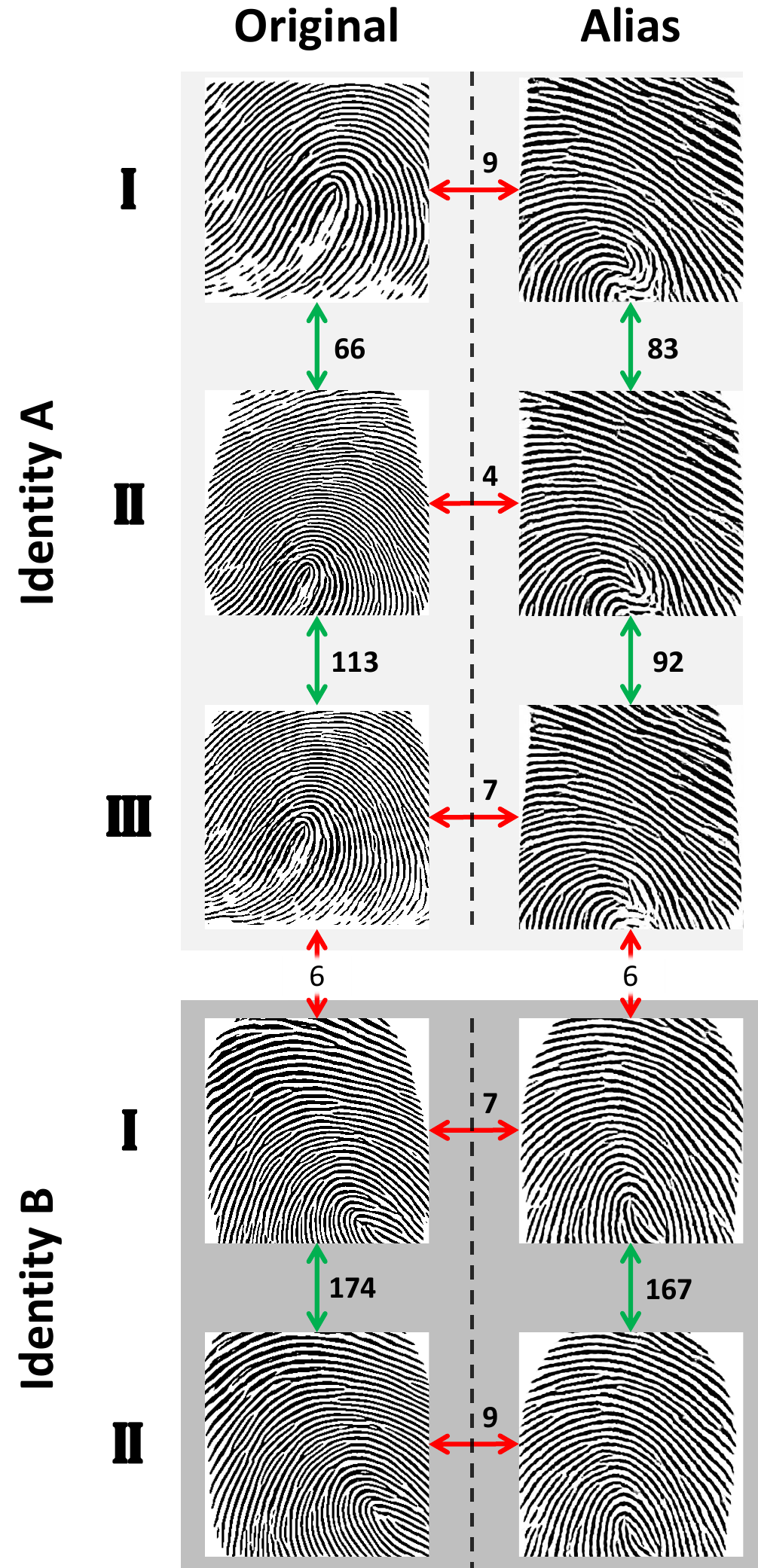}
    \caption{Original impressions (left) match one another, and \mbox{ProxyPrints} aliases (right) likewise match among themselves, but no cross–matching occurs. Scores are Bozorth3 similarity values; a score above 40 denotes acceptance. The visualization confirms that the \mbox{ProxyPrints} transform preserves intra-identity similarity while eliminating correlations between the source and alias domains.}
    \label{fig:proxyprint_comparison}
\end{figure}

\section{\mbox{ProxyPrints}}
\label{sec:insight_proxyprints}

\subsection{Key Insight of \mbox{ProxyPrints}}

The core idea behind ``\mbox{ProxyPrints}'' is to replace stored biometric templates with secure, deterministic aliases, akin to how passwords are hashed to mitigate misuse upon exposure. Direct hashing is infeasible for fingerprints, as matching requires tolerance to natural scan variability. Instead, \mbox{ProxyPrints} transforms each fingerprint into a synthetic alias using a generative model seeded with a deterministic key. The key is used to rotate the embedding space of the generative model, producing unique, repeatable aliases per finger. This enables consistent authentication while ensuring that compromised aliases reveal no usable biometric information.

This design shifts the paradigm of biometric security by eliminating the need to store real fingerprint data while using existing off-the-shelf matchers. Each ProxyPrint is generated through a deterministic process keyed with randomness, ensuring consistent mapping for authentication while enabling unlinkability and easy re-issuance if compromised.

Commercial matchers such as NBIS~\cite{nist2010nbis}, VeriFinger~\cite{verifinger2022}, and SourceAFIS~\cite{galuska2015sourceafis} operate as backend SDKs that ingest raw fingerprint images and return similarity scores, independent of the capture hardware.  \mbox{ProxyPrints} supplies its synthetic alias at this same image interface, leaving scanners, databases, and proprietary matching code untouched.

\mbox{ProxyPrints} aligns with Kerckhoffs’s principle \cite{kerckhoffs2023cryptographie}, asserting that security should rely solely on the secrecy of a secret key rather than on the secrecy of the system itself. Consequently, \mbox{ProxyPrints} openly relies on a securely managed key to randomize biometric templates, ensuring robust security while maintaining transparency and operational flexibility.

Practical benefits of \mbox{ProxyPrints} include:

\begin{itemize}
    \item \textbf{One-Time Training:} The encoder and generator are trained once; no retraining is needed for deployment, updates, or key rotation.
    
    \item \textbf{Open Source Compatibility:} The architecture and weights can be fully open-sourced without compromising security, as protection depends solely on key secrecy.
    
    \item \textbf{Seamless Integration:} \mbox{ProxyPrints} can be applied to any 3rd party biometrics system as a transparent plug and play solution compatible with any other fingerprint matcher.
    
\item \textbf{Breach Detection:} If a scanned fingerprint matches an alias stored in the database without having passed through \mbox{ProxyPrints}, it provides \textbf{proof} that this database has been breached. This is because aliases exist only internally and are never externally visible, any direct match against them reveals the use of a stolen or reconstructed print.
    
    \item \textbf{Key Rotation:} Secret keys can be rotated efficiently to produce new identity mappings without retraining the model.
\end{itemize}

\subsection{\mbox{ProxyPrints} Composition}

\subsubsection{Overview}
We define the \mbox{ProxyPrints} transformation as a function $T : \mathcal{X} \rightarrow \mathcal{X}'$, where $\mathcal{X}$ is the space of genuine fingerprint images and $\mathcal{X}'$ is a derived space of alias identities. Given a fingerprint impression $x_p^i \in \mathcal{X}$, the transformation $T(x_p^i) = x_p'^{i} \in \mathcal{X}'$ produces a realistic fingerprint image that retains matcher compatibility while representing a new, consistent identity, determined by a fixed key but unlinkable across keys or transformation domains.

This transformation is modeled as a composition of three functions:
$$
T(x_p) = \text{De}(\text{Align}(\text{En}(x_p))) = x_p'
$$
The encoder $\text{En}$ maps the fingerprint image $x_p$ into a $n$-dimensional embedding $e \in \mathbb{R}^{n}$. The aligner applies a fixed key-dependent rotation $R$ to this embedding, resulting in $e' = R(e)$. Finally, the decoder $\text{De}$ reconstructs a fingerprint-like image $x'_p$ from $e'$.

The transformation $T$ is deterministic, mapping each input to a unique alias, and is forward-only: applying $T$ again produces a new, unrelated alias:
$$
T(T(x_p^i)) = T(x_p'^{i}) = x_p''^{i} \neq x_p^i
$$
As shown in Fig.~\ref{fig:proxyprint_comparison}, \mbox{ProxyPrints} preserves high similarity within each domain, original or alias, while breaking correspondence between them. This enables alias generation that is revocable, unlinkable across keys, and compatible with conventional fingerprint matchers.

\subsubsection{Function Components}
The encoder $\text{En}(x_p) = e$ projects the fingerprint image onto a $n$-dimensional unit hypersphere. This angular encoding ensures that impressions from the same finger cluster closely in embedding space, while those from different fingers remain well-separated.

The aligner then applies a fixed, key-specific rotation $R$, transforming the embedding to $e' = R(e)$. This allows consistent alias generation without retraining the model. Because the rotation is stable and deterministic for a given key, impressions from the same finger will produce embeddings that rotate to the same alias identity.

The decoder $\text{De}(e')$ generates the alias image $x_p'$ from the rotated embedding. Since the encoder is trained for stability under intra-class variation, different impressions $x_p^i$ and $x_p^j$ of the same finger produce embeddings that map to alias fingerprints with high similarity.

We quantify this similarity using a matcher function $\text{match}(\cdot, \cdot)$, which measures the degree of correspondence between two fingerprint images. This matcher reflects how likely it is that two images originate from the same individual. Under this definition, the final transformer $T$ ensures that $\text{match}(T(x_p^i), T(x_p^j))$ is high if $x_p^i, x_p^j$ are impressions of the same finger and $\text{match}(T(x_p^i), T(x_p^k))$ is low if $x_p^k$ is from a different finger.
This matcher-aware aliasing framework enables secure, revocable biometrics while maintaining compatibility with standard recognition pipelines.

\subsubsection{Encoder and Decoder Training}\label{subsec:encdec_training}
To promote diverse and identity-specific mappings through the transformation $T$, we intentionally train the encoder and decoder models \textit{separately}, rather than as a conventional autoencoder. This decoupling encourages the encoder to learn compact, discriminative identity representations, and the decoder to independently learn how to generate high-quality fingerprint images conditioned on those representations. The result is a more expressive transformation space, allowing $T$ to produce more unique and robust aliases.

The encoder $\text{En}$ is trained as a Siamese network using a triplet loss formulation, following the approach from FaceNet \cite{schroff2015facenet}. During training, the network receives a triplet of fingerprint images: two impressions from the same finger and one impression from a different individual. The objective is to minimize the distance between the embeddings of the same identity, while maximizing the distance to the embedding of the different identity. Formally, for a triplet $(x_p^a, x_p^+, x_p^-)$, where $x_p^a$ (anchor) and $x_p^+$ (positive) are from the same identity, and $x_p^-$ (negative) is from a different one, the loss encourages:
$$
\| \text{En}(x_p^a) - \text{En}(x_p^+) \|_2^2 + \alpha < \| \text{En}(x_p^a) - \text{En}(x_p^-) \|_2^2
$$
for a margin $\alpha > 0$. Because the embeddings are normalized to lie on the unit hypersphere, angular distance becomes the natural measure of similarity. As a result, the angle between embeddings directly captures fingerprint identity, with closer angles implying higher identity similarity.

% The decoder $\text{De}$ is a generative model based on StyleGAN \cite{karras2019style}, trained to synthesize realistic fingerprint images. StyleGAN models naturally interpret the input latent code (aka the “style vector”) as an angular signal that controls identity-related features in the generated output. This makes it well-suited to our framework, where the decoder receives embeddings from the encoder (or rotated aliases thereof) and must treat them as continuous identity representations. We train the decoder in an unsupervised manner on a large dataset of fingerprint images, optimizing only for visual fidelity and realism without any identity labels. This ensures that the decoder learns a rich, generalizable fingerprint distribution without bias toward specific users or templates.

The decoder $\text{De}$ is a generative model based on StyleGAN~\cite{karras2019style}, trained to synthesize realistic fingerprint images. In our framework, the input to the decoder is a normalized latent embedding, which lies on the unit hypersphere. This design induces an angular interpretation of identity, where different directions in latent space correspond to distinct identities. Although StyleGAN does not inherently enforce such a structure, our embedding normalization and rotation-based aliasing introduce this geometric property. The decoder is trained in an unsupervised manner on a large dataset of fingerprint images, optimizing solely for visual fidelity and realism. No identity labels are used, ensuring that the model learns a generalizable distribution of fingerprint appearances without overfitting to specific users or templates.

The encoder was trained for 200 epochs with a batch size of 128 using the Adam optimizer (learning rate $5.21 \times 10^{-6}$, weight decay $1 \times 10^{-5}$). We applied random rotation ($\pm7^\circ$) and $11\%$ random cropping for augmentation. A 90/10 train-test split was used, with $20\%$ of test identities reserved for validation. The decoder was trained with the default configuration from the following StyleGAN implementation.\footnote{\url{https://github.com/huangzh13/StyleGAN.pytorch}}.

Together, these two independently trained components form a transformation pipeline $T$ that is highly expressive, match-aware, and identity-consistent, ensuring that similar embeddings generate similar images, while allowing distinct aliases to be derived via controlled embedding manipulations.

\subsubsection{Aligner and Security Properties}
The aligner is the core mechanism that enables \mbox{ProxyPrints} to provide security and revocability. It operates by rotating the encoder’s output embedding using a fixed, secret key, thereby deterministically reassigning each fingerprint to a new identity while preserving intra-identity matchability. This transformation ensures that each deployment generates a unique and isolated alias space that cannot be linked or reversed without the key.

Given a fingerprint image $x_p$, the encoder produces an embedding $e = \text{En}(x_p^i) \in \mathbb{R}^n$, constrained to lie on the unit hypersphere. The aligner then applies a rotation matrix $R \in \text{SO}(n)$, parameterized by $n$ angles $\{\theta_1, \theta_2, \dots, \theta_n\}$, one per embedding axis. This yields a new embedding $e' = R(e)$, which is then decoded to the alias fingerprint image ${x'}_p = \text{De}(e')$.

Through an empirical analysis, we found that rotating a single axis by 10 degrees is sufficient to produce a new identity, as determined by a matcher, a full discussion can be found in the Appendix~\ref{sec:boundary_analysis}. This means that each axis supports roughly 36 distinguishable identity states, yielding a total of $36^n$ possible unique identity mappings across the embedding space. Even for $n = 512$ (the embedding size we use in this paper), this results in a combinatorially vast set of transformations, effectively eliminating any possibility of key guessing or exhaustive alignment.

\begin{figure}
    \centering
    \includegraphics[width=1\linewidth]{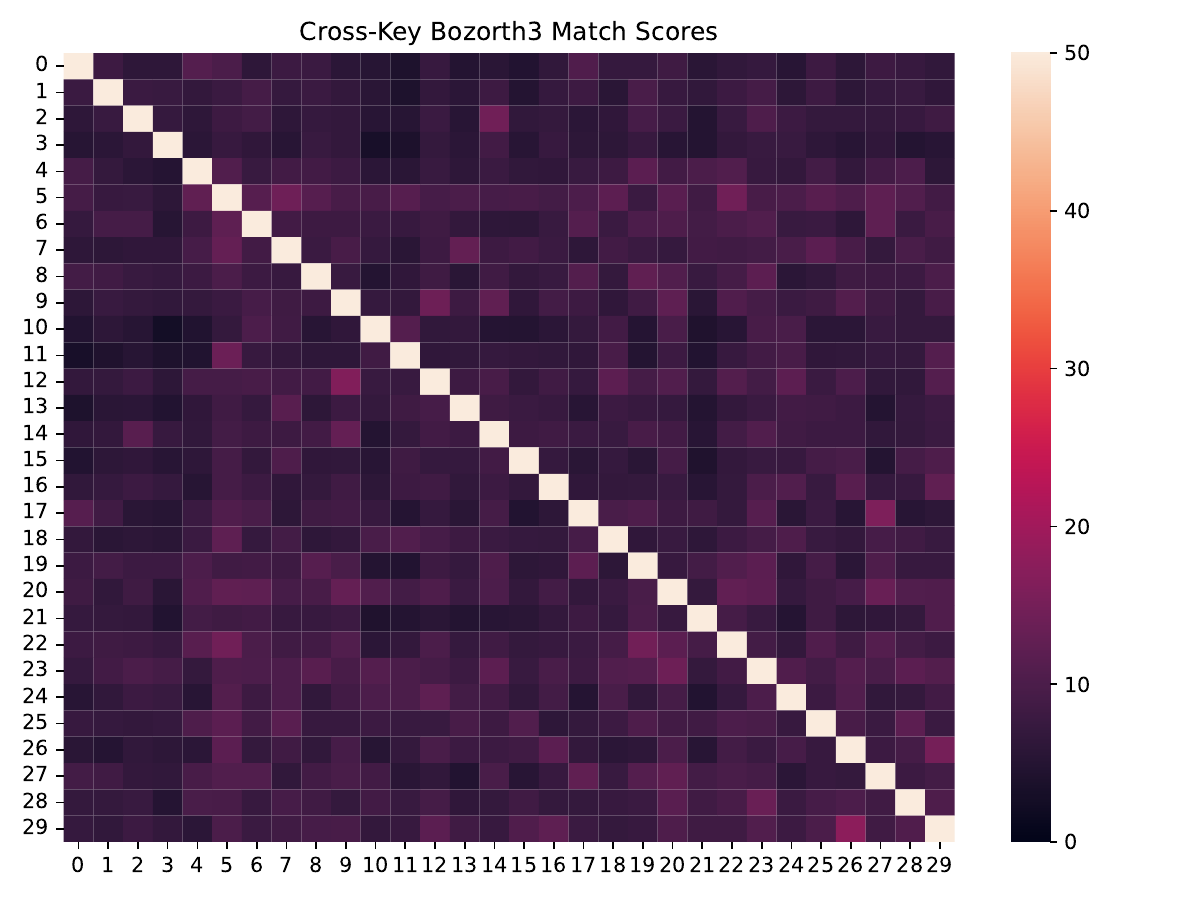}
    \caption{Cross-key similarity heatmap showing average Bozorth3 match scores between \mbox{ProxyPrint} aliases transformed under 30 different keys. High scores appear only on the diagonal (same key), confirming that different keys produce unlinkable aliases below the match threshold of 40.}
    \label{fig:cross_key_heatmap}
\end{figure}

Each key defines a unique rotation of the embedding space, and therefore a unique alias universe. Figure~\ref{fig:cross_key_heatmap} confirms that these identity spaces are disjoint: matcher scores between aliases produced under different keys fall below the acceptance threshold, showing that no identity leakage occurs across deployments.

An attacker who obtains the encoder and decoder weights but not the secret key faces an impractical challenge. The ultimate goal of the adversary is to recover an original fingerprint image corresponding to one of the $M$ enrolled identities. However, without the key, the transformation cannot be reversed. To break the system, the attacker would need to collect the original physical fingerprint of every enrolled user from some external source, and then iterate through candidate rotations until the entire set aligns with stored aliases.

This attack becomes unfeasible for two reasons. First, obtaining one genuine print per enrolled identity is logistically and legally difficult at scale. Second, without knowing the key, aligning even a small subset of embeddings across such a high-dimensional space is computationally prohibitive, especially when the number of embedding dimensions $n$ greatly exceeds the number of users $M$. In this setting, partial knowledge is insufficient; the adversary must solve for the entire rotation using complete ground truth, which makes the attack practically implausible.

While this analysis is geometric rather than cryptographic in nature, the result is a strong practical defense. \mbox{ProxyPrints} does not rely on secrecy of model parameters, and it avoids single-point failure modes like template inversion. Future work may explore formal reductions or hybrid approaches, but the current aligner design already presents a prohibitive barrier to adversarial re-identification in deployed systems.

\begin{figure*}[t]
    \centering
    \includegraphics[width=\linewidth]{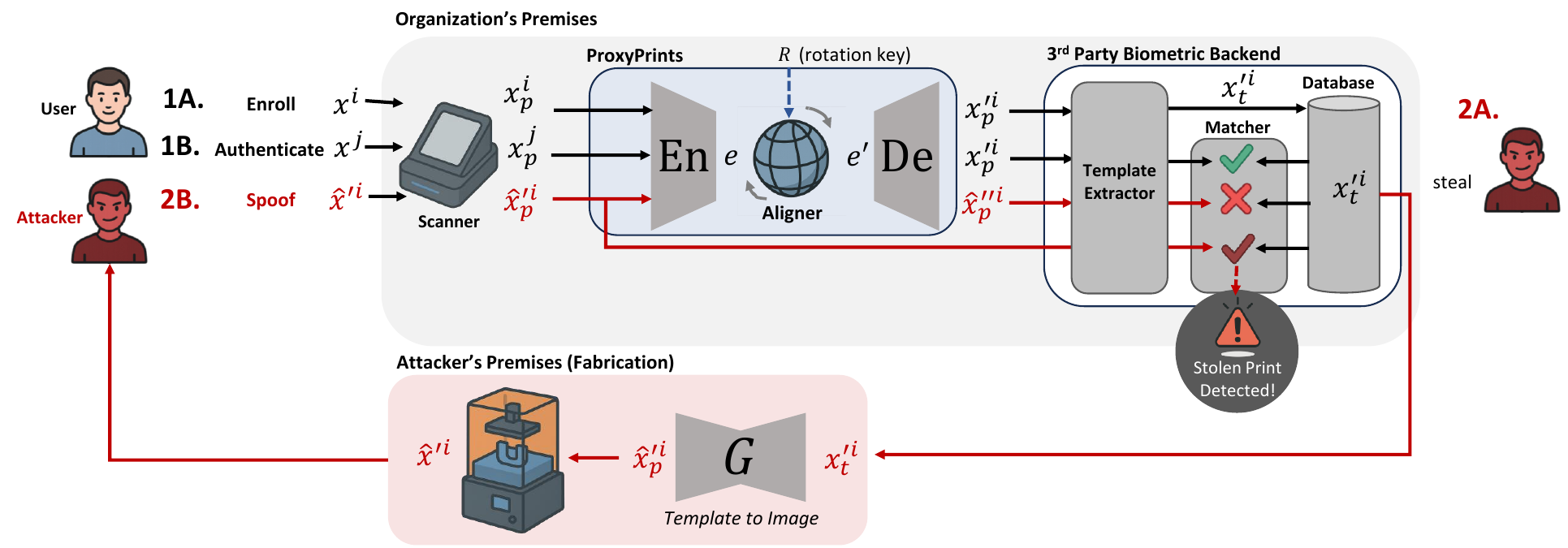}
    \caption{\mbox{ProxyPrints} deployed in a standard fingerprint pipeline. Scenario 1A (enrollment) the user places a live finger on the scanner; \mbox{ProxyPrints} transforms the captured impression into an alias print, and the third-party backend stores the resulting template. Scenario 1B (authentication) a later live impression is transformed into the same alias, so the backend matcher accepts the attempt. Scenario 2A (theft) an adversary compromises the backend and extracts stored alias impressions or templates. Scenario 2B (spoof) the adversary fabricates a replica from the stolen data and submits it to the scanner; \mbox{ProxyPrints} re-aliases the replica, it no longer matches any template, the attempt is blocked, and the system raises a \emph{Stolen Print Detected} alert.}
    \label{fig:proxyprints_architecture}
\end{figure*}

\subsection{\mbox{ProxyPrints} Framework}
The \mbox{ProxyPrints} framework is designed with three primary goals: 
(1) to prevent stolen fingerprint templates from matching legitimate user entries 
(2) to detect spoofing and misuse attempts, protecting biometric systems from reconstruction attacks.
(3) to be a drop-in software addition to existing fingerprint matching systems.

The \mbox{ProxyPrints} middleware inserts transparently between the scanner and any minutiae matcher, so deployment requires no modification of existing matching or registration code. During enrollment every raw scan is first processed by \mbox{ProxyPrints}, deterministically mapped to an alias, and only that alias (or extracted template) is written to the database. 
% Because the matcher never encounters genuine templates, a record stolen from the database cannot be replayed: when it re-enters through the scanner it is flagged as stolen, triggering an alert to the sysadmin for key-rotation.
As illustrated in Fig.~\ref{fig:proxyprints_architecture}, the middleware ensures that any template stolen from the database fails to re-authenticate because a second pass through the aliasing function produces a different identity, and the mismatch raises an immediate alert to the administrator.

A single key governs the rotational mapping applied in embedding space. The production key resides in a hardware security module alongside other sensitive credentials. Upon suspected compromise the administrator can nullify the stored aliases and generate a new random rotation key and then re-enroll the user's fingerprints. This is similar to the standard practice of requiring all users to update their passwords after a credential database breach. Key rotation is a lightweight orthogonal shift in embedding space, so it requires no retraining of the generative model and completes within seconds while preserving separability for legitimate users. This capability to revoke and reissue biometric identities converts fingerprints from irrevocable secrets into renewable credentials, closing a long-standing gap in practical cancellable biometrics.

\subsubsection{Compatibility and Latency}
\mbox{ProxyPrints} works regardless of how the 3rd party biometric software ultimately stored the scanned fingerprint image -as a template or image. We found that our \mbox{ProxyPrints} module only adds a latency of 200ms per fingerprint. This ensures \mbox{ProxyPrints} remains compatible with real-time and high-throughput biometric systems.

\section{Evaluation}
\label{sec:evaluation}
We evaluate the \mbox{ProxyPrints} framework across three key dimensions. First, we assess whether applying \mbox{ProxyPrints} preserves recognition performance by measuring its impact on existing matcher accuracy. Next, we evaluate the system’s ability to resist spoofed prints stolen from the system's database. In other words, that applying the transformation twice yields a distinct identity: $T(T(x)) \neq T(x)$. Finally, we test \mbox{ProxyPrints} capacity for post-rejection attribution: when a stolen print is submitted and rejected, we evaluate how well the system can prove that \textit{this} database has been compromised, that the used print originated from this organization.

\subsection{Experiment Setup}

\noindent\textbf{Dataset.}
We evaluate \mbox{ProxyPrints} using the publicly available \textbf{LivDet} datasets from 2009, 2011, 2013, and 2015 \cite{galbally2009livdet2009, yambay2012livdet2011, ghiani2013livdet, mura2015livdet2015}, which together comprise 3,516 distinct identities and a total of 24,700 fingerprint impressions. Because the number of impressions per identity varies, we perform random resampling to balance the dataset, ensuring exactly 20 impressions per identity.

Given the relatively small size of this dataset by modern deep learning standards (due to limited public access to large-scale fingerprint databases) we augment the data by horizontally flipping all fingerprint images. This simple transformation allows us to generate synthetic ``mirror'' identities, effectively doubling the number of distinct identity representations. After augmentation, the final dataset contains approximately 126,000 impressions.

To avoid identity leakage and ensure fair evaluation, we split the dataset by identity (not by impression) into a training set of 3,160 identities and a held-out test set of 356 identities.

\noindent\textbf{\mbox{ProxyPrints} Setup.}
As described in Section~\ref{sec:method}, we train the encoder and decoder models independently using only the training portion of the dataset. In all experiments, the encoder outputs a 512-dimensional embedding ($n = 512$). The complete training pipeline and code for \mbox{ProxyPrints} will be made available as an artifact of this paper.\footnote{https://github.com/PenlessPan/ProxyPrints}

\noindent\textbf{Metrics.}
We evaluate performance using the industry-standard Bozorth3 matcher \cite{ko2007user}. To quantify system accuracy, we report common biometric metrics, including area under the receiver operating characteristic curve (ROC AUC), area under the precision-recall curve (PRAUC), and Equal Error Rate (EER), which reflects the point where false acceptance and false rejection rates are equal. Higher AUC values and lower EER indicate better performance.

We also report threshold-based classification metrics at the default matcher threshold of 40, including Accuracy, Precision, Recall, and F1-score. Finally, to provide insight into system behavior at specific operating points, we report accuracy at a fixed false acceptance rate (FAR = 0.1) and a fixed false rejection rate (FRR = 0.1).

\begin{table}
\centering
\caption{Matcher Performance Comparison With and Without \mbox{ProxyPrints} Middleware}\label{tab:matcher-performance}
\begin{tabular}{l|cc}
\toprule
\textbf{Metric} & \textbf{Bozorth3} & \textbf{\mbox{ProxyPrints} + Bozorth3} \\
\midrule
ROC AUC & 0.93 & 0.86 \\
PRAUC & 0.95 & 0.87 \\
EER Rate & 0.14 & 0.19 \\
Accuracy (T=40) & 0.83 & 0.75 \\
Precision (T=40) & 1.00 & 0.89 \\
Recall (T=40) & 0.66 & 0.58 \\
F1 (T=40) & 0.79 & 0.70 \\
\midrule
Accuracy@FAR=0.1 & 0.87 & 0.78 \\
Accuracy@FRR=0.1 & 0.83 & 0.75 \\ 
\bottomrule
\end{tabular}
\end{table}

\begin{figure}[t]
    \centering
    \includegraphics[width=1\linewidth]{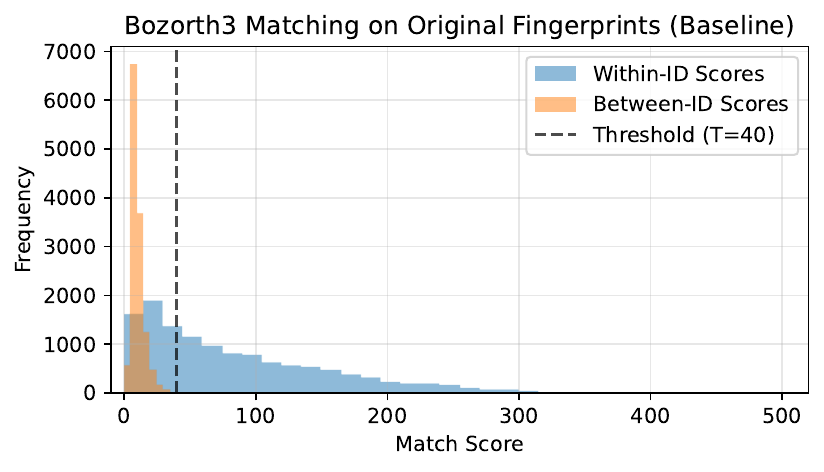}
    \includegraphics[width=1\linewidth]{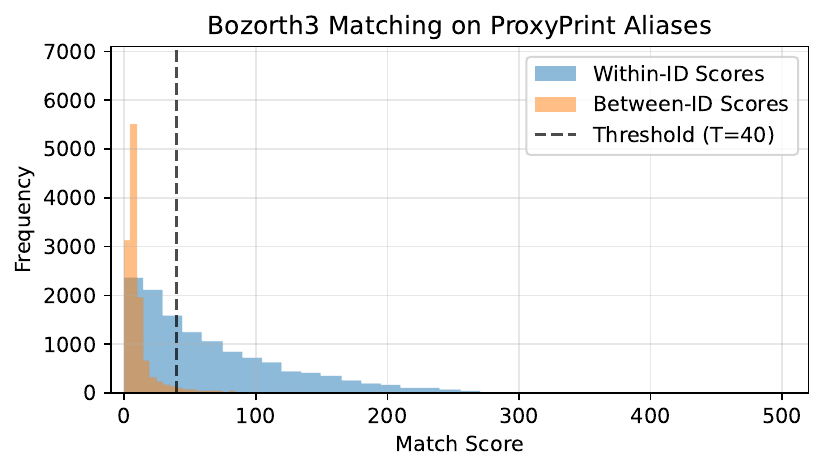}
    \caption{Score distributions before (top) and after (bottom) integration of \mbox{ProxyPrints}. The baseline shows Bozorth3 match scores without transformation, while the defense distribution reflects scores after \mbox{ProxyPrints} aliasing.}
    \label{fig:score-distributions}
\end{figure}

\subsection{Matcher Performance}
We begin by evaluating whether \mbox{ProxyPrints} affects genuine user authentication. Table~\ref{tab:matcher-performance} compares baseline Bozorth3 performance against the \mbox{ProxyPrints}-augmented pipeline across standard biometric metrics. Results show modest reductions in AUC (0.93 to 0.86) and an accuracy at threshold 40 of 0.83 to 0.75, but overall performance remains within acceptable operational bounds. 
Figure~\ref{fig:score-distributions} further illustrates this point by showing the distribution of Bozorth3 match scores before and after aliasing. After applying \mbox{ProxyPrints}, within-identity scores decrease \textit{slightly} due to the variability introduced by alias generation, while between-identity scores remain concentrated near zero. This shift reflects the cost of transformation but still preserves a usable separation between genuine and impostor scores.

\begin{figure}
    \centering
    \includegraphics[width=1\linewidth]{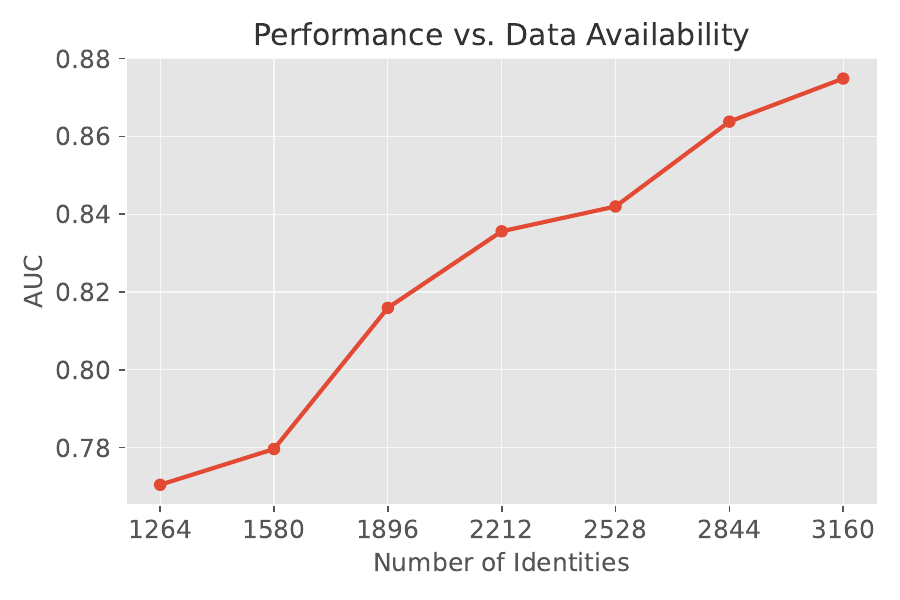}
    \caption{Performance scales with data availability: as the number of identity points increases, the AUC of \mbox{ProxyPrints} steadily improves and converges towards the baseline performance of 0.93.}
    \label{fig:learning_curve}
\end{figure}

We attribute the modest decline in matching performance \textbf{primarily to the limited training data} available for our generative aliasing model. As illustrated in Figure~\ref{fig:learning_curve}, performance improves \textit{consistently} as the number of training identities increases, indicating that the performance gap between \mbox{ProxyPrints} and the baseline narrows with more data. This trend strongly suggests that the observed degradation is not intrinsic to the \mbox{ProxyPrints} framework, but rather a consequence of data scarcity. In practice, any organization with access to a moderately sized fingerprint dataset, larger than those publicly available, should be able to adopt \mbox{ProxyPrints} with minimal or no performance trade-off.

In summary, these findings confirm that \mbox{ProxyPrints} can preserve core matcher functionality, supporting its integration into existing systems without significant degradation of genuine match accuracy.

% \subsection{Detecting Adversary Usage}

% % Stolen Fingerprint Images Experiment
% % tested on the entire database (356 identities with multiple impressions)
% % it's both tested aas "what happens if they steal the fingerprint and try to authenticate with it we detect it really well.

% We evaluate \mbox{ProxyPrints}' efficacy in both blocking unauthorized template usage and detecting biometric spoofing attempts.

% To measure \mbox{ProxyPrints}' robustness against attacks utilizing stolen fingerprint aliases, we inject \mbox{ProxyPrints} into the matcher pipeline and assess how many synthetic attempts successfully match stored templates. 

\begin{figure}[t]
    \centering
    \includegraphics[width=1\linewidth]{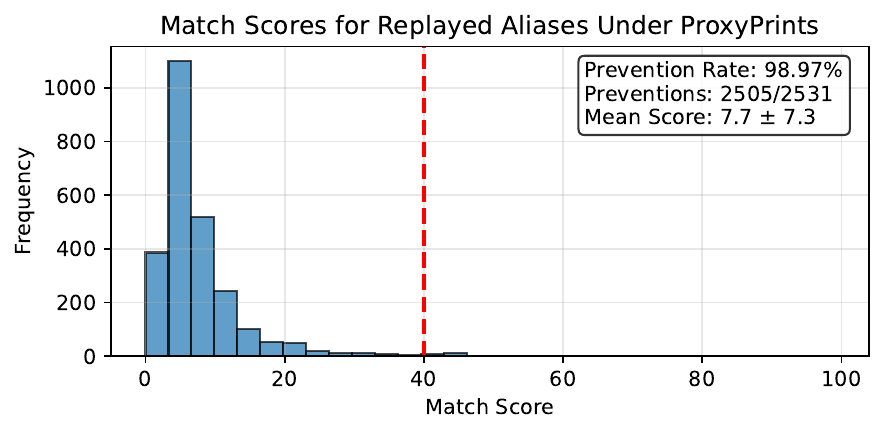}
    \caption{\mbox{ProxyPrints}' performance in detecting stolen fingerprints (aliases) since they undergo a second transformation (i.e., $T(T(x))$). The figure plots the distribution of match scores, with the vertical dashed line indicating the decision threshold. 
}
    \label{fig:stolen-spoofs}
\end{figure}

\subsection{Rejecting Stolen Spoofs}
We now evaluate how effectively \mbox{ProxyPrints} defends against attackers who attempt to authenticate using stolen biometric data, specifically, by replaying a stored alias fingerprint image. In this threat model, an adversary gains access to an alias stored in the system’s database and mistakenly assumes it represents a valid physical fingerprint. They then submit this stolen alias in an attempt to gain access.

To simulate this attack, we install \mbox{ProxyPrints} into the fingerprint matcher pipeline and pass the stolen alias through the full transformation process once again. Since \mbox{ProxyPrints} applies a deterministic transformation to any input, reprocessing the stolen alias results in a completely new fingerprint image: $T(T(x)) \neq T(x)$. Thus, the second alias should no longer match the original one stored in the system.

Figure~\ref{fig:stolen-spoofs} illustrates the result of this evaluation. The figure shows the distribution of match scores between stored aliases and their doubly-transformed counterparts, with the vertical dashed line indicating the system’s decision threshold. The match scores are clearly separated from the genuine distribution, with nearly all falling well below the threshold, yielding a detection rate of 98.97\%.

This result confirms that \mbox{ProxyPrints} is highly effective at blocking replay attacks involving stolen aliases. By design, even if an adversary acquires a stored alias, reusing it will yield a non-matching identity, thus preserving the integrity and security of the authentication system.

\subsection{Breach Detection}
Beyond preventing unauthorized access, \mbox{ProxyPrints} also supports reliable breach detection. Since alias fingerprints are generated and used entirely within the system, and never exposed externally, any attempt to match an external input directly against a stored alias indicates that an internal alias has likely been compromised.

This detection is triggered when a rejected input impression $x$ matches a stored alias $x'$ in the database \textbf{without} being transformed via the \mbox{ProxyPrints} pipeline, i.e., without applying $T(x)$. Such a match suggests that the input is a stolen alias originating from the system itself. When detected, the system can raise alerts, supporting forensic analysis and enabling timely key rotation to contain the breach.

\begin{figure}[t]
    \centering
    \includegraphics[width=1\linewidth]{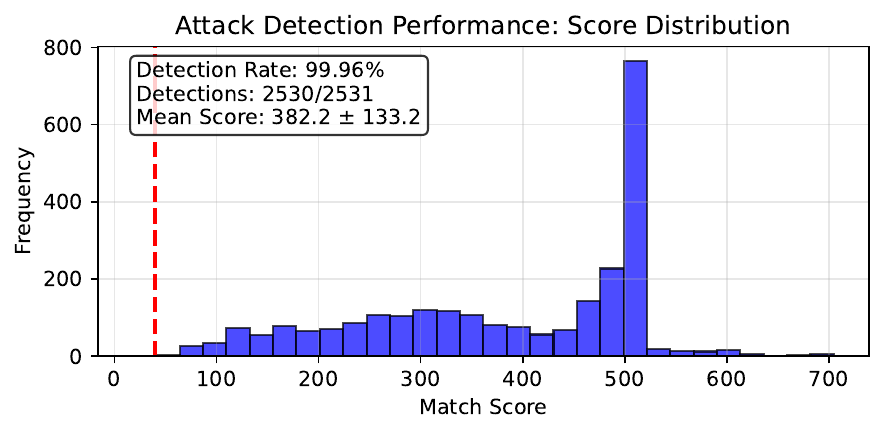}
    \caption{Distribution of match scores for synthetic replay attempts processed through \mbox{ProxyPrints}. A sharp separation occurs at the decision threshold (red line), enabling a 99.96\% detection rate (2530/2531).}
    \label{fig:breach-detection}
\end{figure}

Figure~\ref{fig:breach-detection} presents the distribution of match scores between stolen aliases and the corresponding entries in the database. \mbox{ProxyPrints} achieves a detection rate of 99.96\%, successfully flagging 2,530 out of 2,531 breach attempts. This high accuracy provides strong, actionable evidence that a specific alias from the internal system has been reused externally, offering administrators clear confirmation of a compromise and the means to respond swiftly.

\section{Conclusion}
In conclusion, \mbox{ProxyPrints} provides a practical and scalable defense against the growing threat of fingerprint template inversion and physical spoofing attacks. By introducing a transparent, matcher-agnostic aliasing mechanism, \mbox{ProxyPrints} enables revocable, unlinkable biometric identities that preserve authentication accuracy without requiring changes to existing infrastructure. Through rigorous validation, including real-world spoof fabrication and comprehensive performance evaluation, we demonstrate that \mbox{ProxyPrints} effectively prevents replay attacks, supports breach detection, and offers robust security protections. This work marks a significant step toward making cancellable biometrics a deployable reality for modern fingerprint recognition systems.

\section*{Acknowledgments}
This work was funded by the Center for Cyber, Law and Policy (CCLP) at Haifa University and the Israel National Cyber Directorate. Work was also supported by the Zuckerman STEM Leadership Program.
The authors would also like to thank Yazan Abbas and Luai Okasha for their help in developing the fingerprint research tool used to help streamline the production of silicone molds and general results.

% TODO: Include specific quantitative highlights from experimental results in conclusion once fully finalized.

\bibliographystyle{IEEEtran}
\bibliography{paper}

% @keren, here add intro text to the objective of the experiment (we want to know how many unique mappingt here are with our aligner) so with n=512 we imperically mesaured how many degree rotation per axis required to change ID (get a mathcer score above 40. we repeated this for all n=512 axis. We found that..
% conclusion: ...therefore we estimate that there \textit{at least} 36^512 possible mappings which is large enough to make manual reversal of \mbox{ProxyPrints} iimpractical as discussed int he paper.

\appendices
\section{Identity Boundary Analysis in Latent Space}
\label{sec:boundary_analysis}
\subsection{Experiment Setup and Goal}

The objective of this experiment is to empirically estimate the local structure and granularity of identity boundaries within specific 2D subspaces of the StyleGAN-based fingerprint generator's latent space. Specifically, we aim to determine how identity transitions occur when systematically perturbing the latent space along controlled 2D rotation paths.

To achieve this, we define 2D planes in the latent space by selecting pairs of consecutive dimensions and generate sparse unit-norm embeddings by rotating a vector in each plane across the full $[0^\circ, 360^\circ]$ range. We tested 30 consecutive planes using dimension pairs $(0,1), (1,2), \ldots, (29,30)$ from the 512-dimensional latent space.

For each plane, we create sparse embeddings where only two dimensions are non-zero, defined as:
\begin{flushleft}
\begin{tabular}{rlrlrl}
(1) & $\text{embedding}_{\mathrm{dim}_1}$ &=& $\cos(\theta)$ & & \\
(2) & $\text{embedding}_{\mathrm{dim}_2}$ &=& $\sin(\theta)$ & & \\
(3) & $\text{embedding}_{i}$             &=& $0$ & $\forall i \notin \{\mathrm{dim}_1, \mathrm{dim}_2\}$ \\
\end{tabular}
\end{flushleft}

where $\theta$ ranges from $0^\circ$ to $360^\circ$. This approach isolates the effect of rotations within specific 2D subspaces while maintaining unit vector magnitude.

For each rotation angle, we synthesize a fingerprint using the StyleGAN generator and extract its minutiae using NIST's \texttt{mindtct}. The generated fingerprints are then compared using \texttt{Bozorth3} to evaluate identity similarity against a reference fingerprint generated at $0^\circ$ in the same plane.

We define an identity change as occurring when the Bozorth3 similarity score between the reference fingerprint and a rotated fingerprint falls below 40. Using a systematic scanning approach followed by binary search refinement, we record the last angle before a detectable identity change occurs. This provides a high-resolution measure of how identity boundaries are structured within each 2D subspace.

By repeating this process across 30 different planes, we empirically map local identity transition patterns and gain insight into the anisotropy and resolution at which identity representations change in specific directions of the latent space. 

\subsection{Results and Observations}

\begin{figure*}[t]
    \centering
    \begin{minipage}[t]{0.49\linewidth}
        \centering
        \includegraphics[width=\linewidth]{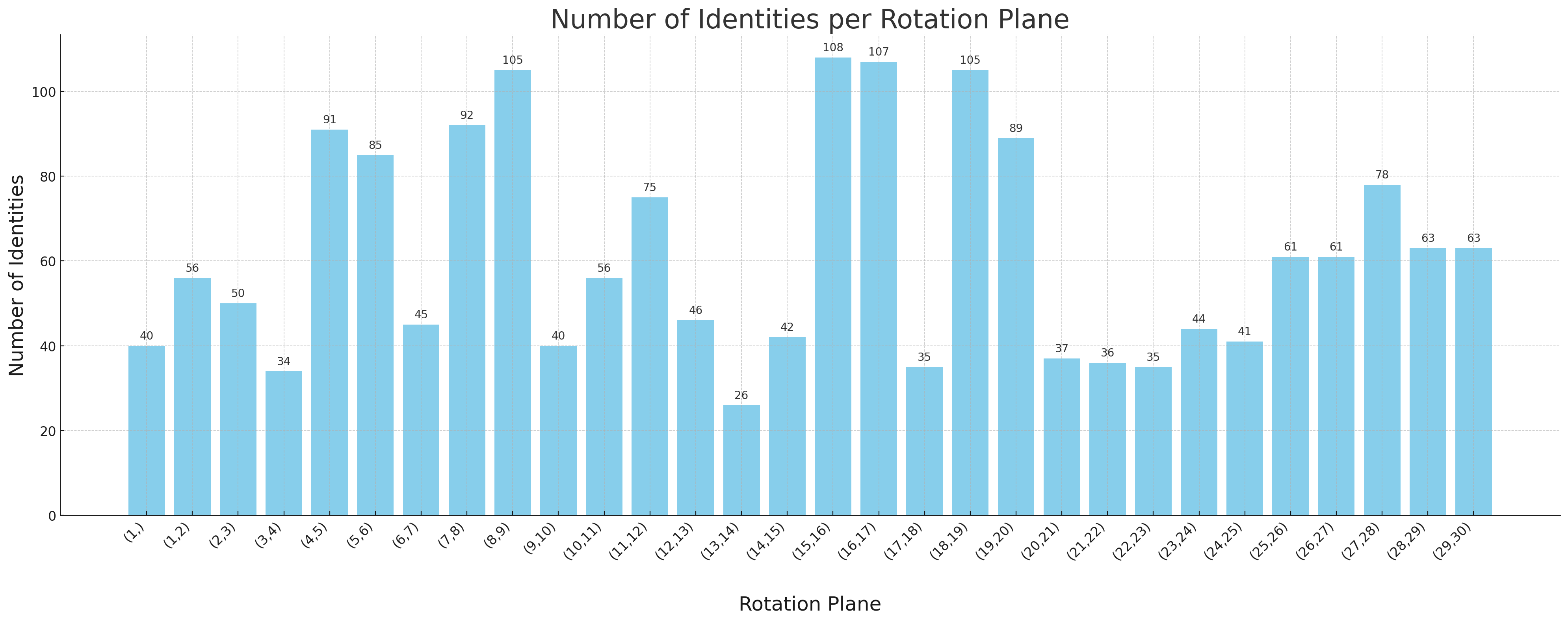}
        \caption{Number of unique identities detected per axis pair over a 360° rotation.}
        \label{fig:identity_count}
    \end{minipage}
    \hfill
    \begin{minipage}[t]{0.49\linewidth}
        \centering
        \includegraphics[width=\linewidth]{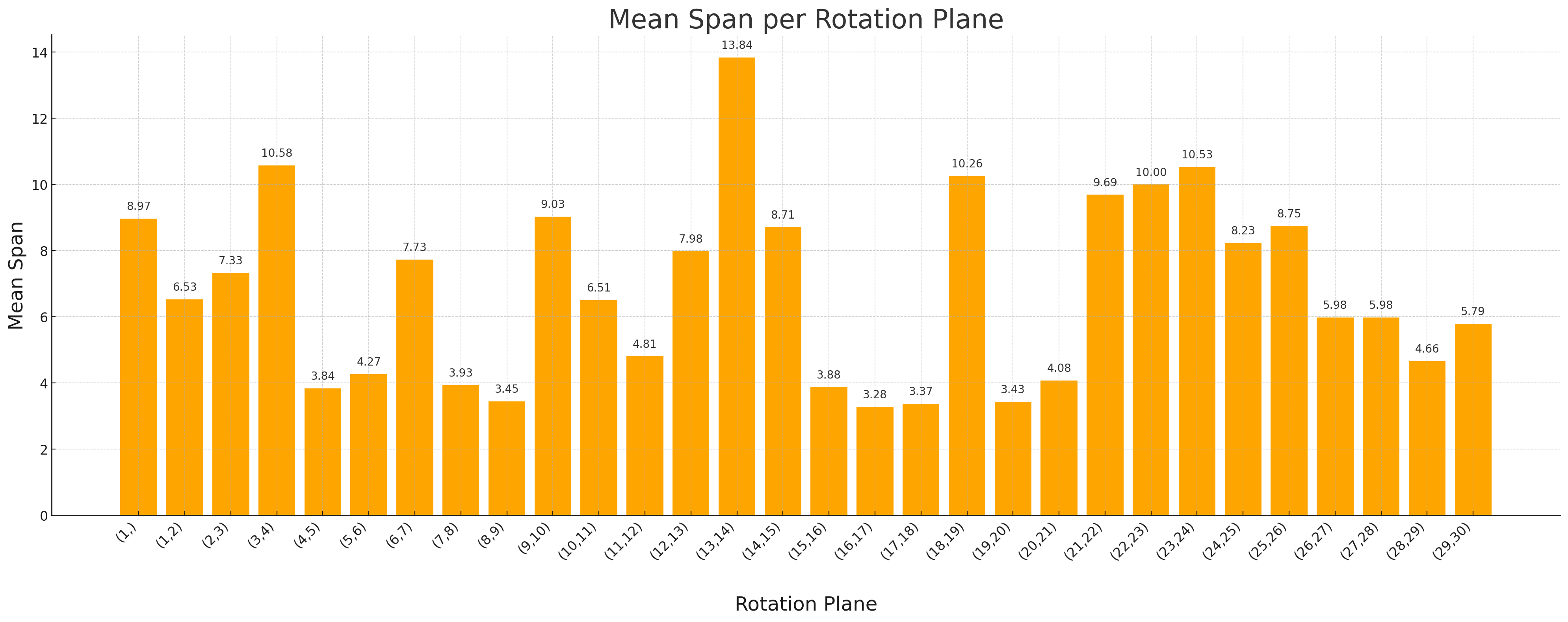}
        \caption{Mean angular span per identity for each axis pair. The orange bars reflect the average degrees each identity spans before a transition occurs.}
        \label{fig:mean_span}
    \end{minipage}
\end{figure*}

Figure~\ref{fig:identity_count} shows the number of unique identities detected for each axis pair, using a similarity threshold of 40 with Bozorth3. We observed significant variation across axis pairs. Some planes, such as (15,16), (16,17), and (19,20), revealed over 100 unique identities, suggesting high identity resolution in those subspaces. In contrast, some other axis pairs, such as (13,14), exhibited relatively few identities, indicating lower sensitivity to changes.

In Figure~\ref{fig:mean_span}, we plot the average angular span (in degrees) that each identity occupies before a change is detected. This orange bar chart highlights the granularity of identity space. Smaller spans suggest that even slight perturbations in that plane result in a new identity. For example, axis pair (13,14) shows a particularly small average span of only $\sim$3.3°, implying very dense identity transitions.

\begin{figure*}
    \centering
    \includegraphics[width=\linewidth]{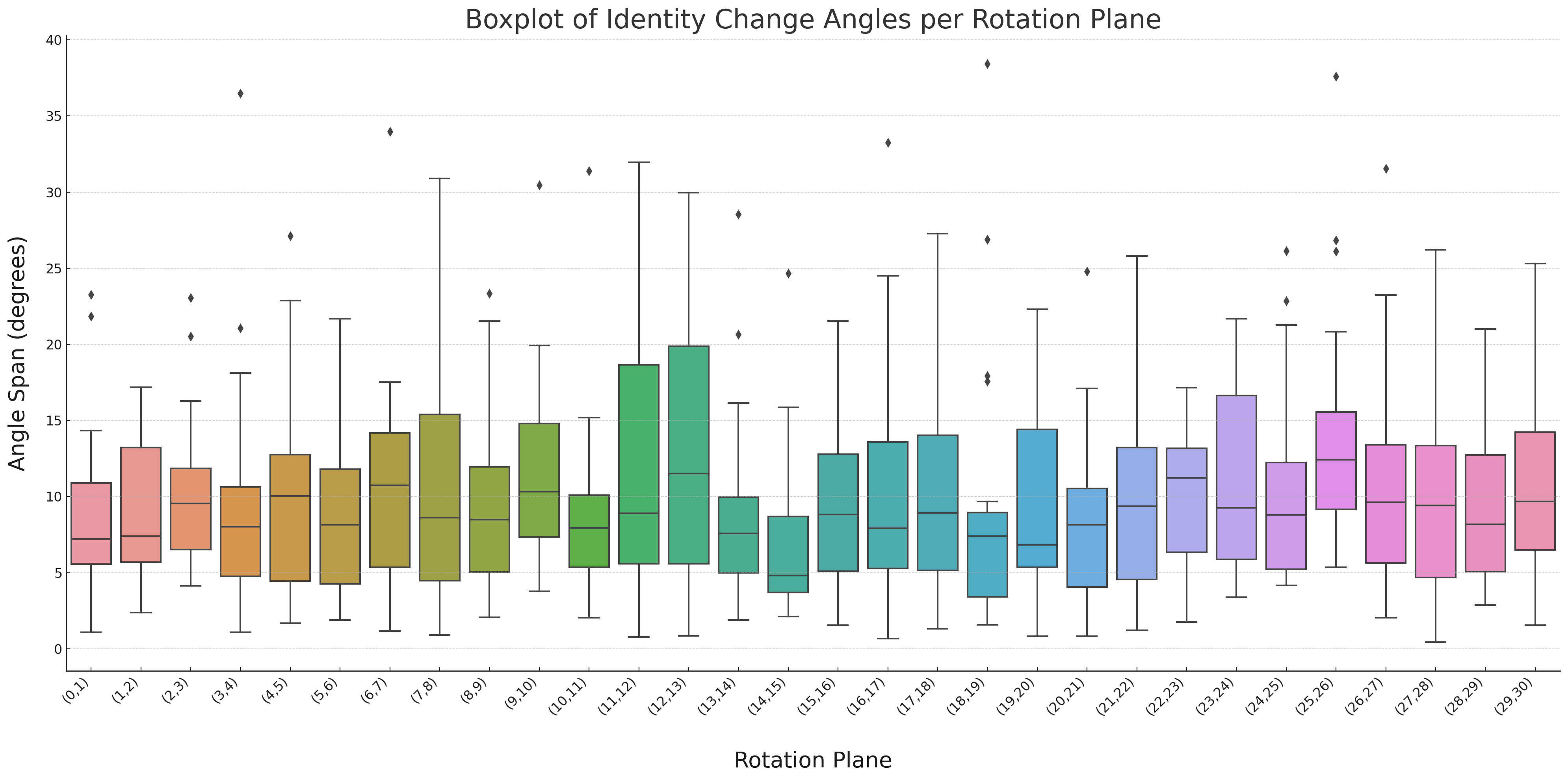}
    \caption{Boxplot of identity change angles per axis. Each box represents the distribution of angular spans between identity changes for a given axis pair.}
    \label{fig:boxplot_angle_span}
\end{figure*}

Figure~\ref{fig:boxplot_angle_span} presents the distribution of angular spans per axis pair in a boxplot format. It reveals that identity stability varies notably across latent planes. Median angular spans range from \(6^\circ\) to \(14^\circ\), indicating differing rates of identity change. Some planes, such as 11--12 and 13--14, exhibit outliers above \(30^\circ\), reflecting unusually persistent identities. While planes like 0--1 and 10--11 show consistent identity spans with tight distributions, others (e.g., 23--24, 24--25) display greater variability. These trends highlight the anisotropic nature of the latent space, where identity sensitivity depends on the direction of traversal.

\textbf{Conclusion:} Based on the average number of identity transitions observed per plane, we conservatively estimate that there are \textit{at least} $36^{512}$ possible identity mappings within the latent space. This combinatorial complexity renders any manual or brute-force reversal of proxy prints infeasible, as discussed in the paper.

\end{document}